\documentclass{elsarticle}

\usepackage{amsmath,amssymb}
\usepackage{latexsym,graphicx,epsfig}
\usepackage{curves,eepic}
\def\lb{\linebreak[4]}
\newcommand{\be}{\begin{equation}}
\newcommand{\ee}{\end{equation}}
\newcommand{\bes}{\begin{subequations}}
\newcommand{\ees}{\end{subequations}}
\newcommand{\bea}{\begin{eqnarray}}
\newcommand{\eea}{\end{eqnarray}}
\newcommand{\bear}{\begin{equation}\begin{array}}
\newcommand{\eear}[1]{\end{array}\label{#1}\end{equation}}
\def\ba{$$\begin{array}}
\def\ea{\end{array}$$}
\def\bra{$\begin{array}}
 \def\era{\end{array}$}

\newcommand{\fr}[2]{\dfrac{{ #1}}{{ #2}}}

\newcommand{\pa}{\partial}
\newcommand{\la}{\langle}
\newcommand{\ra}{\rangle}
\newcommand{\fn}[1]{\footnote{{\sf #1}}}
\def\vak{{\varkappa}}

\newsavebox{\fmbox}

{\end{list}}
\newcounter{enumct}

\newcommand{\bu}{$\bullet$\ }

\usepackage{wrapfig}

\begin{document}
\renewcommand{\tilde}{\widetilde}

\title{Necessity of mixed kinetic term in the description of general system with identical scalar fields}
\author{I. F. Ginzburg
\\
{\it Sobolev Institute of Mathematics and Novosibirsk State
University},\\
{\it Novosibirsk, Russia}}

\begin{abstract}
Most general renormalizable interaction in the system with a set
of scalar fields having identical quantum numbers generates naturally mixed kinetic terms in the Lagrangian. Taking into account these terms leads to modification both the renormalization group equations and the tree level analysis as compare with many published results.
We obtain conditions for non-appearance of such a running
mixing in some important cases.

\end{abstract}

\maketitle

PACS {11.10Gh,11.10.Hi,12.60.Fr}\\

\section{Introduction}\label{secintro}

In this paper I consider systems containing two or more scalar fields with identical quantum numbers. Such systems were studied in numerous papers (e.g. \cite{many}-\cite{dema}). Below I consider a phenomenon that often appears and usually overlooked in analysis of such systems, sometimes -- reasonably, often -- unreasonably. The essence of the phenomenon is the necessity to take into account for renormalizability the mixed kinetic term in the Lagrangian. In fact, this statement can be found in ref.~\cite{Pilaf}\fn{Such term is included in some descriptions for the system of vector gauge fields \cite{SU1}}. However the analysis there does not contain the condition when this term is not necessary or diagonalizing mixing angle is not running. Moreover, during 12 years after publication of \cite{Pilaf} this effect is overlooked as before in many papers (for recent references see e.g. \cite{haber09}, \cite{MaM}).

It is well known that the field mixing in these theories can be naturally accounted by transition to a new basis in $\{\phi_i\}$-space (general rotation plus renormalization) in which both kinetic and mass terms of the Lagrangian become diagonal, and physical particles become different from those described by basic fields. Below we consider {\it diagonalizing mixing angles}, realizing diagonalization of only kinetic term. From the naive glance, the price for the transition to this form is only in the redefinition of coupling constants in potential. However, the situation is more complex.

In the majority of cases a system, containing fields with
identical quantum numbers, originates from the system having some
extra symmetry at very small distances; at these distances our
fields either are members of some more general multiplet or have
quantum numbers, separating these fields. In this region the
discussed field mixing either is forbidden or can be considered as
some generalized rotation with non-running mixing angles. This
symmetry is violated at some intermediate scale either by new
(large distance) interactions or spontaneously. It results in
appearance of field mixing at large distances. That is the {\it
natural theory} in which, going to smaller distances, the effect
of mentioned symmetry violation vanishes, system restores a
primary symmetry globally, the diagonalizing mixing angles become
constant. The key point of our discussion below is the
classification of models with respect to their correspondence to
this {\it naturalness}. We concentrate our discussion on kinetic
mixing only and do not consider well known problems of
diagonalization of physical states (near mass shell) -- see
e.g.~\cite{Pilaf} for details.

In the sect.~2 we describe considered models and show how mixed kinetic term appears. In the sect.~3 we describe some specific features of renormalization procedure for the system with fields having identical quantum numbers. The sect.~4 is devoted to classification of possible Lagrangians in respect of specific discrete $Z_2$ symmetry which appears in correspondence with classification in respect of mentioned naturalness of theory. In the sect.~5 we found condition for this naturalness -- condition for non-appearance of running kinetic term. In the sect.~6 the corresponding renormalization group equations (RGE) in one--loop approximation are constructed. In the sect.~7 we consider main physical consequences of the result.

\section{Models}\label{secsimplest}

I consider below two models. The first one is the simplest model containing two real scalar fields $\phi_1$ and $\phi_2$ \eqref{bassclagr0} (see e.g. \cite{many}-\cite{IFG77}). It allows to present all calculations without complex combinatorics. The essence of a phenomenon is clearly seen in this simplest model.
In parallel we consider general Two Higgs Doublet Model (2HDM) which can pretend for description of physical reality.

\paragraph{\bf Simplest model} The "most general" renormalizable Lagrangian of the system with two real scalar fields is written usually as (notation is chosen to be closer to the standard 2HDM case)
\bear{c}
{\cal{L}_\phi}= T_{0\phi} -V ,\;\;
 T_{0\phi}=
 \fr{\pa_\mu \phi_1\;\pa_\mu \phi_1 + \pa_\mu \phi_2\;\pa_\mu \phi_2}{2},\\[3mm] V=V_2+V_4,\;\;
 V_2= \fr{ m_{11}^2\phi_1^2+m_{22}^2\phi_2^2+2m_{12}^2\phi_1\phi_2}{2},\\[2mm]
V_4=
\fr{\lambda_1}{2}\phi_1^4+
\fr{\lambda_2}{2}\phi_2^4+\lambda_3\phi_1^2\phi_2^2
+\lambda_6\phi_1^3\phi_2+\lambda_7\phi_1\phi_2^3
 \,. \eear{bassclagr0}

\paragraph{\bf Two Higgs Doublet Model (2HDM)} One of the most important realizations of the discussed type systems
presents the simplest extension of the minimal SM -- the 2HDM, describing the system of two complex isodoublet scalar fields $\phi_1$ and $\phi_2$ with identical hypercharges (see e.g. \cite{2HDM}-\cite{Ivan1}) with the Lagrangian
 \bear{c}
{\cal{L}}= T_{0\phi} -V ,\;\;
 T_{0\phi}=
 \fr{\pa_\mu \phi_1^\dagger\;\pa_\mu \phi_1 + \pa_\mu \phi_2^\dagger\;\pa_\mu \phi_2}{2},\\[3mm]
V\!=\!-\fr{m_{11}^2\phi_1^\dagger\phi_1\!+\!
 m_{22}^2\phi_2^\dagger\phi_2 \!+\!\left( m_{12}^2 \phi_1^\dagger\phi_2\!+\!h.c.\right)}{2}\,+\!\\[2mm]
+\fr{\lambda_1(\phi_1^\dagger\phi_1)^2 + \lambda_2(\phi_2^\dagger\phi_2)^2}{2} +
\lambda_3(\phi_1^\dagger\phi_1)(\phi_2^\dagger\phi_2) +
\lambda_4(\phi_1^\dagger\phi_2)(\phi_2^\dagger\phi_1) +\\[3mm]+ \left[\fr{\lambda_5(\phi_1^\dagger\phi_2)^2}{2}
+{\lambda_6(\phi_1^\dagger\phi_1)(\phi_1^\dagger\phi_2) + \lambda_7 (\phi_2^\dagger\phi_2)(\phi_1^\dagger\phi_2)} + h.c.\right] .
 \eear{2HDMLagr}

Here the vacuum state with $\la\phi_i\ra\neq 0$ violates EW
symmetry with transition to the representation of the Lagrangian
via physical Higgs fields and Goldstone fields. The states of this
physical Higgs representation are obtained from incident EW
isodoublets by simple decomposition and rotation. This procedure
does not influence the ultra-violet behavior. Therefore, one can
consider our ultra-violet kinetic mixing problem in the primary EW
symmetric basis with the Lagrangian e.g. in the form
\eqref{2HDMLagr}, just as it was done at obtaining positivity
constraints for the potential (see e.g. \cite{dema}).

\paragraph{\bf The reparameterization symmetry} The potential \eqref{bassclagr0}
depends on 8 real free parameters: $m_{11}^2, m_{22}^2$,
$m_{12}^2$ $\lambda_1, \lambda_2, \lambda_3, \lambda_6$,
$\lambda_7$ while potential \eqref{2HDMLagr} depends on 14 free
parameters: real $m_{11}^2, m_{22}^2$, $\lambda_1, \lambda_2,
\lambda_3, \lambda_4$ and complex $m_{12}^2, \lambda_7, \lambda_6,
\lambda_7$.

However, the models contains two fields with identical quantum
numbers. Therefore, they can be described both in terms of fields
$\phi_k$ $(k=1,2)$, used in (\ref{bassclagr0}), \eqref{2HDMLagr},
and in terms of fields $\phi'_k$ obtained from $\phi_k$ by a
global unitary {\it reparametrization transformation} ${\cal F}$
-- rotation in $(\phi_1,\,\phi_2)$ space: \bes\label{rotat}
\bear{c}
 \begin{pmatrix}\phi'_1\\\phi'_2\end{pmatrix}={\cal F}
 \begin{pmatrix}\phi_1\\\phi_2\end{pmatrix}\,
 \Rightarrow\; \begin{pmatrix}\lambda'_i\\m'^2_{ij}\end{pmatrix}=
 {\cal U_F}\begin{pmatrix}\lambda_i\\m^2_{ij}\end{pmatrix} \,.
  \eear{rotats}
 \bear{cl}
 {\cal F}=\begin{pmatrix}\cos\theta &\sin\theta\\-\sin\theta&
 \cos\theta\end{pmatrix}&for \;\; model\;\; (1);\\[4mm]
  \hat{\cal F}=e^{-i\rho_0}\begin{pmatrix}
\cos\theta\,e^{i\rho/2}&\sin\theta\,e^{i(\tau-\rho/2)}\\
-\sin\theta\,e^{-i(\tau-\rho/2)}&\cos\theta\,e^{-i\rho/2}
\end{pmatrix}&for \;\; model\;\; (2).
 \eear{reparam}\ees
The Lagrangians, which can be obtained from each other by these
transformations, describe identical physical reality. They
represent the {\it reparameterization equivalent family of the
Lagrangians}. These transformations form {\it reparameterization
symmetry groups}, studied for 2HDM in detail, e.g. in\cite{Ivan1}.
In particular, physical observables in the simplest model depend
on 7 free parameters, while in 2HDM they depend on 11 free
parameters (cf. discussion in \cite{GinKraw}, \cite{Ivan1}).

\paragraph{\bf Appearance of mixed kinetic term} Let us consider renormalization of two-point Green functions. The diagrams with two external legs appear at two-loop
\begin{wrapfigure}[6]{r}{0.4\textwidth}\vspace{-3mm}
\includegraphics[height=1.1cm,width=0.37\textwidth]{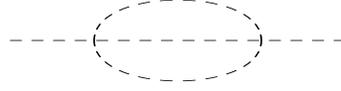}
\caption{\it 2-loop polarization operator}
  \label{propagdiagr}
\end{wrapfigure}
level, see Fig.~\ref{propagdiagr}. These diagrams are
diver\-gent quadratically and describe polarization operators $\Pi_{ab}$. One can choose subtraction point so that its divergent part have form ($a,\,b=1,\,2$)
 \bes\label{2pointz2}\be
 \Pi_{ab}=[(Z_{3,ab}-\delta_{ab})k^2\phi_a\phi_b +(Z_{m,ab}-m_{ab}^2)\phi_a\phi_b]/2 \label{renorm2gen}
\ee
with\fn{These equations are written for the simplest model, the equations for 2HDM have the same structure but slightly more complex form.}
 \bear{c}
Z_{3,11}-1\propto 12\lambda_1^2+4\lambda_3^2+9\lambda_6^2+3\lambda_7^2, \\[2mm] Z_{3,22}-1\propto 12\lambda_2^2+4\lambda_3^2+3\lambda_6^2+9\lambda_7^2,\\[2mm]
Z_{3,12} \propto
\lambda_1\lambda_6 + \lambda_6\lambda_3 + \lambda_3\lambda_7 +
\lambda_7\lambda_2.
 \eear{2pointz21}\ees
Corresponding counter-terms in the Lagrangian
 \be
(Z_{3,aa}-1)\pa_\mu\phi_a\pa_\mu\phi_a/2\;\; \mbox{ and }\;\;
(Z_{m,aa}-m_{aa}^2)\phi_a^2/2\label{CTsim}
  \ee
have the same structure as some terms in the unperturbed Lagrangian and describe renormalization of wave functions and masses. The counter-terms
\be
Z_{3,12}\pa_\mu\phi_a\pa_\mu\phi_1\phi_2\;\; \mbox{ and }\;\;
(Z_{m,12}-m_{12}^2)\phi_1\phi_2\label{CTsimmix}
  \ee
describe the kinetic and mass field mixing respectively.

The appearance of counter-term with $Z_{3,12}$ means that {\bf the Lagrangians \eqref{bassclagr0}, \eqref{2HDMLagr} in the general case are non-renormalizable, for renormalizability the kinetic term $T_{0\phi}$ must be enlarged by the mixed kinetic term}\fn{This phenomenon is similar in some respects to necessity to supplement Yukawa interaction by $\phi^4$ term in the renormalizable Lagrangian, see e.g. \cite{BSh}.},
 \be
 T_{0\phi}\!\!\to\!\! T_\phi\!=\!
 \left\{\!\!\begin{array}{cl}
 \fr{\pa_\mu \phi_1\pa_\mu \phi_1+\pa_\mu \phi_2\pa_\mu \phi_2
 +2\vak \pa_\mu \phi_1{\pa_\mu } \phi_2}{2} &\!for\;(1);\\[3mm]
 \! \fr{\pa_\mu \phi_1^\dagger\pa_\mu \phi_1\!+\!\pa_\mu \phi_2^\dagger\pa_\mu \phi_2\!
 +\!\vak \pa_\mu \phi_1^\dagger{\pa_\mu } \phi_2\!+\!\vak^* \pa_\mu \phi_1{\pa_\mu } \phi_2^\dagger}{2}&\!for\;(2).
 \end{array}\right.\label{kapterm}\ee

\section{Renormalization procedure, case with field mixing}\label{secrenproc}

Let us describe in detail the renormalization procedure for the case when field mixing is possible.

We reduce each step of perturbation theory to the standard construction of $S$-matrix, in which the kinetic term have form $T_{0\phi}$, and together with diagonal mass terms (with $m_{11}^2$ and $m_{22}^2$) it forms the unperturbed Lagrangian ${\cal{L}}_0$,
while other terms (including mixed mass term $\propto m_{12}^2$) are treated as perturbations\fn{In this approach the partial summation of perturbation theory series in $m_{12}^2$ allows to diagonalize mass term. We do not discuss such procedure below.}.

\bu {\bf Procedures $\pmb{\hat{\cal F}_B}$, $\pmb{\hat{\cal Z}_B}$.} Let we have basically the Lagrangian ${\cal L}_B$ with non-diagonal kinetic term $T_\phi$ \eqref{kapterm}. The first stage contains reparameterization rotation $\hat{\cal F}_B$ of form \eqref{rotat}, diagonalizing kinetic term, and subsequent renormalization $\hat{\cal Z}_B$, normalizing all items of kinetic term, with resulting kinetic term $T_{0\phi}$
 \be
{\cal L}_{Bd0}=\hat{\cal Z}_B\hat{\cal F}_B{\cal L}_B.\label{ZFoper}
 \ee

\bu {\bf Procedure $\pmb{\hat{\cal P}_1}$.} The calculation of
radiative corrections in the first nontrivial order gives
counter-terms of the Lagrangian, which lead to the renormalized
Lagrangian
 \be
{\cal L}_B^{R1}=\hat{\cal P}_1{\cal L}_{Bd0}=\hat{\cal P}_1\hat{\cal Z}_B\hat{\cal F}_B{\cal L}_B.\label{ren1oper}
 \ee

\bu To start calculation of next order radiative corrections,
this Lagrangian ${\cal L}_B^{R1}$ is transformed with procedures $\hat{\cal F}_1$ and $\hat{\cal Z}_1$, similar to $\hat{\cal F}_B$ and $\hat{\cal Z}_B$, in order to obtain the renormalized Lagrangian with the same kinetic term $T_{0\phi}$, as in beginning:
 \be
{\cal L}_{Rd1}=\hat{\cal Z}_1\hat{\cal F}_1{\cal L}_B^{R1}. \label{ren1diag}
 \ee

\bu The subsequent iterations are described in the same manner with operators $\hat{\cal P}_2$, $\hat{\cal F}_2$ and $\hat{\cal Z}_2$, etc. The new (but almost trivial) point is an appearance of new diagonalizing procedure ${\cal F}_i$ in each order of perturbation theory. Generally, {\it diagonalizing mixing angles $\theta_i$} in these $\hat{\cal F}_i$ are different --- we obtain {\it running mixing angles} -- an additional subject for renormalization, similar to coupling constants. This very phenomenon is described by introduction of mixed kinetic term in the Lagrangian, with running coefficient.

\section{Different opportunities}

Now we consider different opportunities with respect to {\bf $\pmb
Z_2$ symmetry}, i.~e. invariance of the Lagrangian under
transformations, which prohibit $\phi_1\leftrightarrow\phi_2$
transitions:
 \bear{cl}
 \phi_i^2\to \phi_i^2, \qquad \phi_{1} \phi_{2}\to - \phi_{1} \phi_{2}& for\;\; model\;\;(1);\\[2mm]
 \phi_{i,a}^\dagger \phi_{i,b}\to \phi_{i,a}^\dagger \phi_{i,b}, \qquad \phi_{1,a}^\dagger \phi_{2,b}\to - \phi_{1,a}^\dagger \phi_{2,b}& for\;\; model\;\;(2)\eear{Z2inv}
($a$ and $b$ are weak isospin indices)

({\bf A}) {\it {\bf The dimension 4 ({\em dim4Zs}) $ Z_2$ symmetry} is realized for the operator dimension 4 part of the Lagrangian}, it takes place for the dim4Zs potential with
 \be
{\cal L}_Q:\qquad\lambda_6=\lambda_7=0,\quad \vak=0. \label{dim4Zs}
 \ee

(A1) If additionally $m_{12}^2=0$, we deals with the case of precise $Z_2$ symmetry (for the entire Lagrangian). In this case field mixing is absent in the bare Lagrangians \eqref{bassclagr0}, \eqref{2HDMLagr}. The counter-terms which mix fields $\phi_a$ does not appear.

(A2) If $m_{12}^2\neq 0$, field mixing is obliged by only operator of dimension 2, transitions $\phi_1\leftrightarrow\phi_2$ are allowed near mass shell (at large distances) and forbidden far from mass shell (at small distances) -- {\bf softly violated $\pmb{Z_2}$ symmetry}.

The mixed term $m_{12}^2\phi_1\phi_2$ in the basic potential generates mixed polarization operator $\Pi_{12}$. However, since single $Z_2$ violating term $m_{12}^2\phi_1\phi_2$ has operator dimension 2 (not 4!), it results in the polarization operator only logarithmic divergences, i.e. $Z_{3,12}=0$. Therefore, in this case the mixed kinetic counter-term does not appear. In other words, the discussed mixing takes place at large distances only, it disappear at small distances (large virtualities), that is {\it the natural theory}, discussed above.

The same reasons show that the amplitudes for transitions
$\phi_1\phi_2\to \phi_1\phi_1$, etc., appeared in perturbations,
are convergent and disappear at small distances (at large
virtualities). Hence, they do not give counter-terms like
$\lambda_6\phi_1^3\phi_2$, etc. Therefore, the Lagrangian
\eqref{bassclagr0},\eqref{dim4Zs} is completely renormalizable.

In terms  of sect.~\ref{secrenproc}, in these cases the
diagonalizing mixing operators ${\cal F}_i\equiv 1$ in each order
of perturbation theory.

({\bf B}) {\bf Hidden dimension 4 ${Z_2}$ symmetric ({\em
hdim4Zs}) case } can be obtained from the Lagrangian with dimension 4 $Z_2$ symmetry
\eqref{bassclagr0},\eqref{dim4Zs} with the reparametrization
transformation $\hat{\cal F}_H$, it  is reparameterization equivalent to dim4Zs case:
 \be
 {\cal L}_H=\hat{\cal F}_H{\cal L}_Q.\label{hidlag}
 \ee
This Lagrangian has coefficients $\lambda'_i$ and $m'^2_{ij}$,
expressed via primary values $\lambda_i$ and $m^2_{ij}$ with
transformation \eqref{rotat}. In particular, we have
$\lambda'_6\neq 0$, $\lambda'_7\neq 0$. Mixed kinetic term does
not appear in this stage, since our transformation is simple
rotation and basic kinetic term matrix is unitary one, i.e.
diagonalizing operator $\hat{\cal F}_B=1$.

Let us consider below the case $\lambda_1\neq \lambda_2$, for definiteness. In terms of sect.~\ref{secrenproc} the radiatively renormalized Lagrangian
 \be
{\cal L}_H^{R1}=\hat{\cal P}_1{\cal L}_{H}=\hat{\cal P}_1\hat{\cal F}_H{\cal L}_Q\equiv \hat{\cal F}_H\hat{\cal P}_1{\cal L}_Q.\label{ren1hq}
 \ee
In the primary radiatively renormalized dim4Zs Lagrangian
$\hat{\cal P}_1{\cal L}_Q$ the field renormalization constants
were different, $Z_{3,11}\neq Z_{3,22}$. Therefore, after rotation
$\hat{\cal F}_H$ the mixed kinetic counter-term $Z_{3,12}$
appears, producing mixed kinetic term in ${\cal L}_H^{R1}$. By
construction, it is clear, that the diagonalizing operator
$\hat{\cal F}_1=\hat{\cal F}_H^{-1}$. Moreover, it gives
simultaneously the dim4Zs form of the renormalized Lagrangian
${\cal L}_{Rd1}$. Beginning from the second nontrivial order of
perturbation theory we come to the dim4Zs Lagrangian, without the
mixing kinetic term.

({\bf C}) {\it In the general case of {\bf true hardly violated $Z_2$ symmetry} we have $ {\lambda_6\neq 0}$ and (or) ${\lambda_7\neq 0}$ and the rotation \eqref{rotat}, transforming the Lagrangian to the dim4Zs form \eqref{dim4Zs} does not exist.} In our classification, that is {\it unnatural theory}.

In this case in addition to diagonal terms with $Z_{3,aa}$ the
off-diagonal term $Z_{3,12}k^2\phi_1\phi_2$ appears in the
divergent part of polarization operator. It produces counter-terms
$Z_{3,12}\pa_\mu\phi_1\pa_\mu\phi_2$ in the Lagrangian, which were
absent in the bare Lagrangian ${\cal L}_0$. It means that our
theory with the kinetic term $T_{0\phi}$ is not renormalizable. If
rotation \eqref{rotat}, transforming this potential to the dim4Zs
form, does not exist, {\bf to restore renormalizability, the basic
Lagrangian must be supplemented by the mixed kinetic term}, i.e.
the kinetic term of the Lagrangians \eqref{bassclagr0},
\eqref{2HDMLagr} must be rewritten in the form \eqref{kapterm}. As
a result, the diagonalizing mixing angle is different in different
orders of perturbation theory. Therefore, it is running (see
discussion after \eqref{ren1diag}).

This enlargement of the kinetic term adds new degree of freedom in
the tree level phenomenological analysis and makes more complex
RGE. In many respects the coefficient $\vak$ \ can be treated as
some new coupling constant like $\lambda_i$ (see in more detail RG
equations below).

\section{The condition for non-appearance  of  running mixed kinetic term }

In accordance with previous discussion, the question in the title
can be rewritten in such a form: {\it In what case the coefficient
at mixed kinetic term is non-running, or how one can know whether
the considered potential is true hardly ${Z_2}$ violating one or
we deal with hdim4Zs case}? To find the answer, let us remind that
the $\lambda'_{1-7}$ coefficients of "rotated" hdim4Zs potential
are obtained from parameters of the primary dim4Zs potential
\eqref{dim4Zs} with transformation \eqref{rotat}. Therefore,
coefficients of hdim4Zs potential obey some relations. To obtain
these relations one can express $\lambda'_i$ via $\lambda_i$ and
rotation angles from ref.~\eqref{rotat}. Next, one should consider
these equations as equation like $X[\cos\theta]=0$ and find
condition, at which the form $X$ is so degenerate as
$X[\cos\theta]=0$ at arbitrary $\cos\theta$.

For the 2HDM an elegant form of this condition is obtained with
the aid of geometrical approach of \cite{Ivan1}. In this approach
it is useful to write the general potential \eqref{2HDMLagr} via
irreducible representations of $SU(2)\times U(1)$
reparameterization symmetry group. In this way we obtain tensor
and vector forms in the parameter space, constructed from
coefficients of the Lagrangian,
 \bear{c}
 a_{ij}=\fr{1}{2}\begin{pmatrix} Re\lambda_5 -a &Im\lambda_5&Re(\lambda_6-\lambda_7)\\
 Im\lambda_5&- Re\lambda_5 -a& Im (\lambda_6-\lambda_7)\\
 Re(\lambda_6-\lambda_7)&Im (\lambda_6-\lambda_7)&2a
 \end{pmatrix},\\[7mm]
 a=\fr{\lambda_1+\lambda_2-2\lambda_3-2\lambda_4}{6}\,;\\
 b_i=-\fr{1}{\sqrt{2}}\begin{pmatrix}
Re(\lambda_6+\lambda_7)&Im(\lambda_6+\lambda_7)&\fr{\lambda_1-\lambda_2}{2}
\end{pmatrix}\,.
 \eear{tensorIgor}

In the dim4Zs case these forms  some elements of these objects
becomes equal to zero.  Thus, it is easy to check that in this
case the vector $\Phi_k=e_{ijk}a_{ir}b_rb_j=0$, where $e_{ijk}$ is
standard Levy-Civita tensor. This identity conserves at the
reparameterization transformations. Therefore, the representation
independent condition for the non-appearance of running mixed
kinetic term can be written as three conditions\fn{This form of
condition looks more useful than that obtained in \cite{Ivan1} --
{\it vector $b_i$ must be eigenvector of operator $a_{ij}$}.}
 \be
\Phi_k\equiv e_{ijk}a_{ir}b_rb_j=0\qquad (k=1,\,2,\,3)\,.\label{nomix2HDM}
 \ee

The case in which mixed kinetic term does not appear despite the
explicit  violation of dim4 $Z_2$ symmetry is realized at complete
$\phi_1\leftrightarrow\phi_2$ symmetry of $V_4$, i.e. at
 \be
\lambda_1=\lambda_2,\quad \lambda_6=\lambda_7,\quad \lambda_5=\lambda_5^*\,.\label{phisym}
 \ee
In this case the diagonalizing operator ${\cal F}_1$ of
sect.~\ref{secrenproc} is degenerated to the form, proportional to
$\mathbf{1}$, and reason for mixed kinetic term does not appear.
One can check directly that in this case the condition
\eqref{nomix2HDM} is also valid. Therefore, rotation to the
hdim4Zs form is possible.

\bu {\bf Effect of fermions, variations due to Yukawa interaction.}

The systems of scalar fields are considered usually together with
fermion fields $\psi_a$, having Yukawa interaction with both our
scalars. In particular, for the simplest model
 \be
 {\cal{L}}_\psi=
 i\bar{\psi}_a\overleftrightarrow{\hat{\pa}}\psi_a-m_a\bar{\psi}_a\psi_a+
\sum_a (g_{1a}\phi_1+g_2\phi_{2a})\bar{\psi}_a\psi_a\,.\label{fermlagr}
 \ee
With this term $Z_2$ symmetry \eqref{Z2inv} takes place if only
$g_{2a}g_{1a}=0$ for each $a$ --- in addition to $m_{12}=0$,
$\lambda_6=\lambda_7=0$. (For 2HDM, the similar condition looks as
the condition that each right-handed fermion field is coupled to
only one scalar, $\phi_1$ or $\phi_2$. In particular, that are
Model I or Model II in classification of \cite{Hunter}.)

If $g_{1a}g_{2a}\neq 0$, the $Z_2$ symmetry is violated hardly,
giving -- via loop correction -- counter-terms like $\lambda_6$,
$\lambda_7$ and $\vak$. And -- {\it vice versa} -- if initially it
was, e.g. $\lambda_7=0$, $g_{2a}=0$, but $\lambda_6\neq0$ or (and)
$\vak\neq 0$, the loop corrections make necessary to add in the
bare Lagrangian terms with non-zero $\lambda_7$, $g_{2a}$ and
$\vak$  for renormalizability. The condition for the non-running
diagonalizing mixing angles for scalar fields can be written as an
existence of representation of the potential, which has softly
$Z_2$ violated form in scalar sector and simultaneously
$g_{2a}g_{1a}=0$ in Yukawa sector. In particular, for 2HDM this
property is violated in Model III for Yukawa sector (where some
right fermions are coupled to both fields $\phi_1$ and $\phi_2$).

\bu {\bf Three notes}.

{\bf A}. It is easy to find conditions for the non-appearance of mixed kinetic term in two-loop approximation, it follows from \eqref{2pointz2}:
\bear{cl}
\lambda_1\lambda_6 + \lambda_6\lambda_3 + \lambda_3\lambda_7 +
\lambda_7\lambda_2=0\, &for \;\; model (1);\\[2mm]
\lambda_1\lambda_6^* + \lambda_2\lambda_7^* + (\lambda_3+\lambda_4)(\lambda_6^*+\lambda_7)+\lambda_5^*\lambda_6+
\lambda_5\lambda_7^*=0\,&for \;\; model (2)\eear{2-loop}
These conditions are valid only for two-loop approximation. In higher orders these conditions are supplemented by new and new conditions, having no non-trivial solution, except the case of complete $\phi_1\leftrightarrow\phi_2$ symmetry in $V_4$ \eqref{phisym}.

{\bf B}. One can consider \eqref{2HDMLagr} as particular case of
the Lagrangian of theory with 4 complex scalar fields which are
the components of two isospinors $\phi_1$, $\phi_2$. We denote it
here as 4S theory. The coefficients of this 4S Lagrangian are
constrained by condition that this Lagrangian can be written via
isospinors $\phi_i$. These constraints acquire more complex form
after general rotation of all fields in 4S space.

Just as it was discussed above, some terms in the potential of 4S
theory generate mixed kinetic terms with different coefficients
$\vak_{ab}$ ($a\neq b= 1,\,2,\,3,\,4$). General condition for
non-appearance of such terms can be written separately. However in
our case (when the Lagrangian allows the existence of 2HDM form
written via isospinors $\phi_i$) all these conditions are covered
by \eqref{nomix2HDM}.

{\bf C}. The general transformation of the Lagrangian, including
reparameterization and dilations of fields, allows to eliminate
the terms with $\lambda_6$, $\lambda_7$ from potential (like in
the case of dim4Zs). Simultaneously in the kinetic term the
diagonal terms acquire different normalizations, and the mixed
kinetic term appears. This approach is useful in the analysis of
tree approximation and extrema of potential (see \cite{Ivan2}),
but it results in complexities in the study of perturbation
theory. In particular, the counter-terms with $\lambda_6$,
$\lambda_7$ appear in radiative corrections, violating the $Z_2$
symmetry hardly. This representation does not seem to be available
for the study of the discussed problem.

\section{Modified RGE for invariant charges at $k^2\gg |m_{ij}^2|$, the simplest model}

The renormalization group analysis of considered systems was done
e.g. in refs.~\cite{2HDMrg1}, \cite{2HDMrg2}. However, the mixed
kinetic term was not take into account there. Here we describe
main features of the RGE only for invariant charges in the
ultraviolet region for the case of true hard violation of $Z_2$
symmetry, i.e. with the Lagrangian \eqref{bassclagr0},
\eqref{kapterm}. (Equations for propagators, etc. can be obtained
after that by standard methods.)

\begin{wrapfigure}[12]{r}{0.3\textwidth}\vspace{-3mm}
\includegraphics[height=4cm,width=0.25\textwidth]{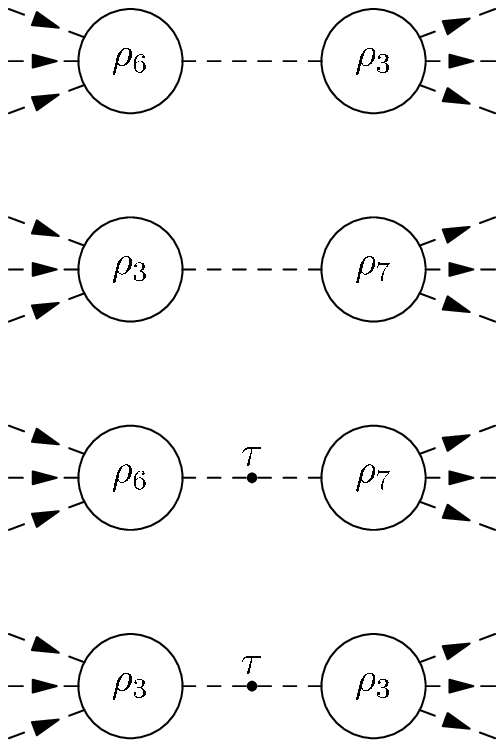}
\caption{} \label{figrgdiagr}
\end{wrapfigure}
The discussion of RGE looks more simple if we add to our system
the interaction with fermions in the form \eqref{fermlagr} with
$g_1g_2\neq 0$, violating $Z_2$ symmetry. The advantage of this
case is that here the mixed kinetic term appears in one-loop
approximation while in the pure scalar case it appears first in
two loops only.

One can imagine two ways for RG analysis of discussed situation:

\bu The kinetic term \eqref{kapterm} is transformed to the
diagonal form by the change of the type of $ \phi_a =A_{ab}\Phi_b$
with $a,\,b=1,\,2$ and suitable $A_{ab}$. After this
transformation the coefficients of potential are changed
$\lambda_i\to \Lambda_i$ but the potential keeps its general form
\eqref{bassclagr0}. In accordance with the previous discussion,
the perturbations produce the scale dependent mixed kinetic term.
Then one must repeat diagonalization (anew at each new scale and
each new iteration).

\bu I prefer to use {\bf another way} in which {\it the mixed kinetic term is treated as an additional contribution into ${\cal
L}_{int}$ with new coupling given by the coefficient at this term}. The diagonalization of the kinetic term can be performed at the final stage.

In this approach, e.g., the typical tree diagrams for the process
$\phi_1\phi_1\phi_2\to\phi_1\phi_2\phi_2$ have the form of
fig.~\ref{figrgdiagr}, where open blob $\rho_i$ corresponds to a full vertex $\rho_i\leftarrow \lambda_i$, dark point corresponds to a full kinetic mixing $\tau\leftarrow\vak$ -- see \eqref{chdef}.

As usual (see \cite{BSh}), we introduce five 1PI 4-scalar-vertexes
$\Delta_{abcd}$ with\lb $a,\,b,\,c,\,d=1,\,2$, two 1PI fermion-scalar vertexes $\Gamma_a$,
nominator of fermion propagator and
matrix numerator of boson
propagator $s$ and $d_{ab}$, defined in the considered region via
complete propagators as $S=\hat{k}s(k^2)/k^2$ and\lb $D=\fr{1}{k^2}\begin{pmatrix} d_{11}&d_{12}\\
                              d_{21}&d_{22}\end{pmatrix}$.
The typical definitions for invariant charges are similar to well
known ones~\cite{BSh}, for example
 \bear{c}
\rho_4=d_{11}^{3/2}d_{22}^{1/2}\Delta_{1112};\;\;
\sigma_1=sd_{11}^{1/2}\Gamma_1;\;\;
\tau_{12}=d_{11}^{1/2}d_{22}^{1/2}d_{12}.
 \eear{chdef}
In the considered simplest case the new invariant charge
$\tau_{12}=\tau_{21}$ and we omit label at $\tau$.

In calculations below we assume, for definiteness,
  $1\gtrsim\vak\gg\lambda_i\sim ï¿-g_i^2 $.

\begin{wrapfigure}[8]{r}{0.4\textwidth}\vspace{-2mm}
\includegraphics[height=2cm,width=0.37\textwidth]{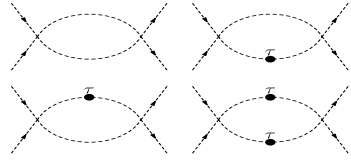}
\caption{\it 4-vertex in one loop}
\label{figvertex}
 \end{wrapfigure}
Now, for example, the typical 4-scalar vertex diagrams
in one loop
approximation have the form of fig.~\ref{figvertex}. Similar
correction must be included in fermion polarization operator.
Corresponding RGE and $\beta$-functions are calculated easily via
known loop integrals with new simple combinatorics. In the one-loop
approximation, for example ($L=\ln(k^2/\mu^2)$),
 \bear{c}
\fr{d \rho_1}{dL}\equiv \beta_{\lambda 1}= 9\rho_1^2+\rho_3^2-4\sigma_1^4+4\sigma_1^2\rho_1+\fr{9}{4}\rho_4^2+\\
+18\vak\rho_1\rho_4+9\vak^2\rho_1\rho_3+\fr{9}{2}\vak^3\rho_4^2\,;\\ \fr{d \sigma_1}{dL}\equiv \beta_{g 1} =\fr{5}{2}\sigma_1^3+\fr{1}{2}\sigma_1\sigma_2^2+\vak \sigma_1^2\sigma_2 \,;\\
\fr{d \tau}{dL} \equiv \beta_\tau = 2\sigma_1\sigma_2 + \vak(\sigma_1^2 + \sigma_2^2) +c\sum\limits_{k=0}^3C_k\vak^k\,.\\[2mm]
C_0=
12(2\rho_1\rho_4 + \rho_4\rho_3 +
\rho_3\rho_5 +
2\rho_5\rho_2)\,,\;\; C_3=16\rho_1\rho_2+4\rho_3^2+10\rho_4\rho_5.
 \eear{rgeqs}
In the sum $\sum C_k\vak^k$ we add two-loop terms, related to the
scalar self-interaction, Fig.~\ref{propagdiagr}, to remind that
the modification of RGE takes place even in the system without
fermions, $c$ is the numerical coefficient and $C_k$ are easily
calculated (we present $C_0$ and $C_3$) but these values are not
interesting for our discussion.

Other equations can be written easily in the similar way.
Combination of results of \cite{KazSh} with simple combinatorics
allows to write the complete system of RGE in two-loop
approximation.

\section{Discussion}

\bu We prove that for the description of the most general system
with a number of scalar fields, having identical quantum numbers,
an additional mixed kinetic term is necessary. The obtained RGE
like \eqref{rgeqs} demonstrate that in the case of hard violation
of $Z_2$ symmetry the fields $\phi_1$ and $\phi_2$ are mixed at
small distances and the mixing parameters (angle and
normalizations) vary with variation of distance (renormalization
scale) $\mu$. In other words, the scalar fields cannot be
separated from each other even at very small distances ({\it
unnatural theory}).

Taking into account this mixed kinetic term makes more complex both RGE and phenomenological analysis.

\bu We find conditions \eqref{nomix2HDM}, at which the coefficient
at this additional kinetic mixing is not running. These conditions
describe models with hdim4Zs potential. In this respect, the
reparameterization equivalent dim4Zs representation of this
potential is preferable for detail studies. In addition, for this
representation also in the Yukawa interaction (for 2HDM) each
right fermion is coupled to only one Higgs boson, i.e. $g_1g_2=0$
for each right fermion field.

If mentioned conditions are valid, the scalar field mixing at
small distances does not vary with the change of distance. Such
theory seem {\it natural} for the description of reality.\\

{\it Acknowledgments}. This paper is an extended version of the
report presented at {\it Dmitry Shirkov fest, {\bf Renormalization
group 2008}, Dubna, September 1, 2008}. It can be considered as
some indirect continuation of my diploma work \cite{Gin56}
prepared under guidance of D.V.~Shirkov. I am very thankful him
for invitation to the activity in this field. I am grateful to
M.~Krawczyk, P.~Osland, G.~Haber for discussions which show me
necessity of this publication. I am also grateful to L. Dias-Cruz,  A. Efremov, I.~Ivanov, F.~Jegherlener, D. Kazakov, K.Kanishev, E. Ma, M. Maniatis, H. Nielsen,  A.~Pilaftsis, V.~Serbo for interesting discussions. I am thankful to
P.~Chankowski, M.~Perez-Victoria for new examples \cite{SU1}. This
paper is supported by grants RFBR 08-02-00334-a and
NSh-1027.2008.2.

\end{document}